\documentclass[aps,prb,twocolumn,floatfix,superscriptaddress,showpacs]{revtex4}
\usepackage[dvips]{graphicx}

\begin{document}
\title{Density functional study of oxygen vacancies at the SnO$_2$ surface and subsurface
sites}
\author{F. \surname{Trani}}
\email{fabio.trani@unina.it}
\homepage[Visit: ]{http://www.nanomat.unina.it}
\affiliation{Coherentia CNR-INFM and Dipartimento di Scienze Fisiche, Universit\`a di Napoli Federico II,
Complesso Universitario Monte S. Angelo, via
Cintia, I-80126 Napoli, Italy}
\author{M. \surname{Caus\`a}}
\email{mauro.causa@unina.it}
\affiliation{Dipartimento di Chimica, Universit\`a di Napoli Federico II,
Complesso Universitario Monte S. Angelo, Via Cintia, I-80126 Napoli, Italy}
\author{D. \surname{Ninno}}
\author{G. \surname{Cantele}}
\affiliation{Coherentia CNR-INFM and Dipartimento di Scienze Fisiche, Universit\`a di Napoli Federico II,
Complesso Universitario Monte S. Angelo, via
Cintia, I-80126 Napoli, Italy}
\author{V. \surname{Barone}}
\affiliation{Dipartimento di Chimica, Universit\`a di Napoli Federico II,
Complesso Universitario Monte S. Angelo, Via Cintia, I-80126 Napoli, Italy}
\affiliation{Istituto per i Processi Chimico-Fisici del CNR, Area della Ricerca, via G. Moruzzi 1, I-56124 Pisa, Italy}
\begin{abstract}

Oxygen vacancies at the SnO$_2 (110)$ and $(101)$ surface and subsurface sites have been studied
in the framework of density functional theory by 
using both all-electron Gaussian and pseudopotential plane-wave methods. The all-electron
calculations have been performed using the B3LYP exchange-correlation
functional with accurate estimations of energy gaps and density of states.
We show that bulk oxygen vacancies are responsible for the appearance of a fully occupied flat energy level lying at about 1 eV above the top valence band, and an empty level resonant with the conduction band. Surface oxygen vacancies strongly modify the surface band structures with the appearance of intragap states covering most of the forbidden energy window, or only a small part of it, depending on the vacancy depth from the surface.
Oxygen vacancies can account for electron affinity variations with respect to the stoichiometric surfaces as well.
A significant support to the present results is found by comparing them to the available experimental data.

\end{abstract}
\maketitle

\section{Introduction}
\label{sec:intro}

Tin dioxide (SnO$_{2}$) is one of the most used and interesting materials for the
development of solid state gas sensors, transparent conductors, and
catalysts.\cite{diebold}
Many physical properties of this and other oxides are driven by
defects which are mostly due to the ease with which the oxygen content can be varied.
In this respect, SnO$_{2}$ surfaces are even more
interesting because the presence of two possible oxidation states of tin (+2 and +4), which are 
combined with the reduced atomic coordination, favors
compositional changes and reconstructions.  Although the
SnO$_{2} (110)$ surface is one of the most studied both theoretically
\cite{oviedo,oviedo1,maki02,asmae} and
experimentally,\cite{diebold,cox,themlin90} its actual structure is still a
matter of debate\cite{diebold} because of the controversial aspects concerning
defects and oxygen adsorption.\cite{gurlo}
Several reconstructions are possible depending on both the preparation conditions and
the sample history. Nevertheless, in a faceted sample, the lowest indices $(110)$ and $(101)$
stoichiometric surfaces seem to be the most likely, to be 
favored from the thermodynamic point of view.\cite{bergermayer}

Bulk and surface oxygen vacancies are extremely important for determining the
electrical conductivity of tin dioxide.\cite{cox} Moreover, they are also
responsible for a very efficient luminescence activity of SnO$_{2}$
nanobelts. First-principles calculations have revealed that bulk oxygen vacancies
and tin interstitials have a low formation energy.\cite{zunger} However,
few theoretical calculations of the energy levels associated to these types of defects
have been published,\cite{maki02,sensato02} probably because standard functionals used in Density Functional Theory (DFT)
fail in giving the correct energy gaps.\cite{parr,becke,becke2,lee, pople}
In this paper, we report on a first-principles study of the electronic and structural properties of SnO$_{2}$ in the presence of oxygen vacancies both in the bulk crystal
and at surface and subsurface sites, focusing on the effects on the energy
levels close to the Fermi energy.
Accurate calculations of the electronic band structure
have been done within an all-electron approach,
which employs the hybrid B3LYP exchange-correlation functional. Electron affinities
have been studied with pseudopotential plane-wave calculations. 

In Sec. \ref{sec:comp}, we outline the computational approaches used in this work.
The structural and electronic properties of defected bulk SnO$_2$ are presented in Sec. \ref{sec:bulk}.
The role of the oxygen vacancies on the surface properties are reported and discussed in the case of the
$(110)$ and $(101)$ surfaces in Secs. \ref{sec:110} and \ref{sec:101}, respectively.
In particular, we consider several defect sites which are both at surface and in subsurface positions, and
analyze their formation energy.
Electron affinity variations driven by defects are studied in Sec. \ref{sec:affinity}.
Finally, in Sec. \ref{sec:concl}, we outline some conclusions.

\section{Computational methods}
\label{sec:comp}

The structural and electronic properties of tin dioxide crystal and surfaces have been
studied, from first-principles, in the framework of DFT.
The first approach is based on an all-electron, Gaussian basis set and is implemented
in the \textsc{crystal\footnotesize{06}} (Ref. \onlinecite{crystal}) package.
The hybrid B3LYP\cite{parr,becke,becke2,lee, pople} approximation for the
exchange-correlation functional has been adopted. This has been shown \cite{muscat,sensato02} to be extremely valid
for band structure calculations, giving reliable
estimations of the energy gaps which are in excellent agreement with
experimental data.
We have verified that this approach
works well for cassiterite (bulk rutile SnO$_2$), as it will be shown below.
The all-electron basis set adopted for tin atoms corresponds to a 3-21G basis set,\cite{dobbs} with the external $sp$ shell being modified with an exponent $\alpha =  0.11$. The basis set adopted for oxygen atoms is the same as that used in Ref. \onlinecite{causa}, where it was applied to magnesium oxide.
For the study of the defected structures, the basis set was enriched by three $sp$ single Gaussian functions (exponents $\alpha=1.0$, $0.30$, and $0.09$) centered at the location of the missing oxygen, in
order to allow the description of trapped electrons.\cite{scorza}

\begin{figure}
\includegraphics[width=.4\textwidth]{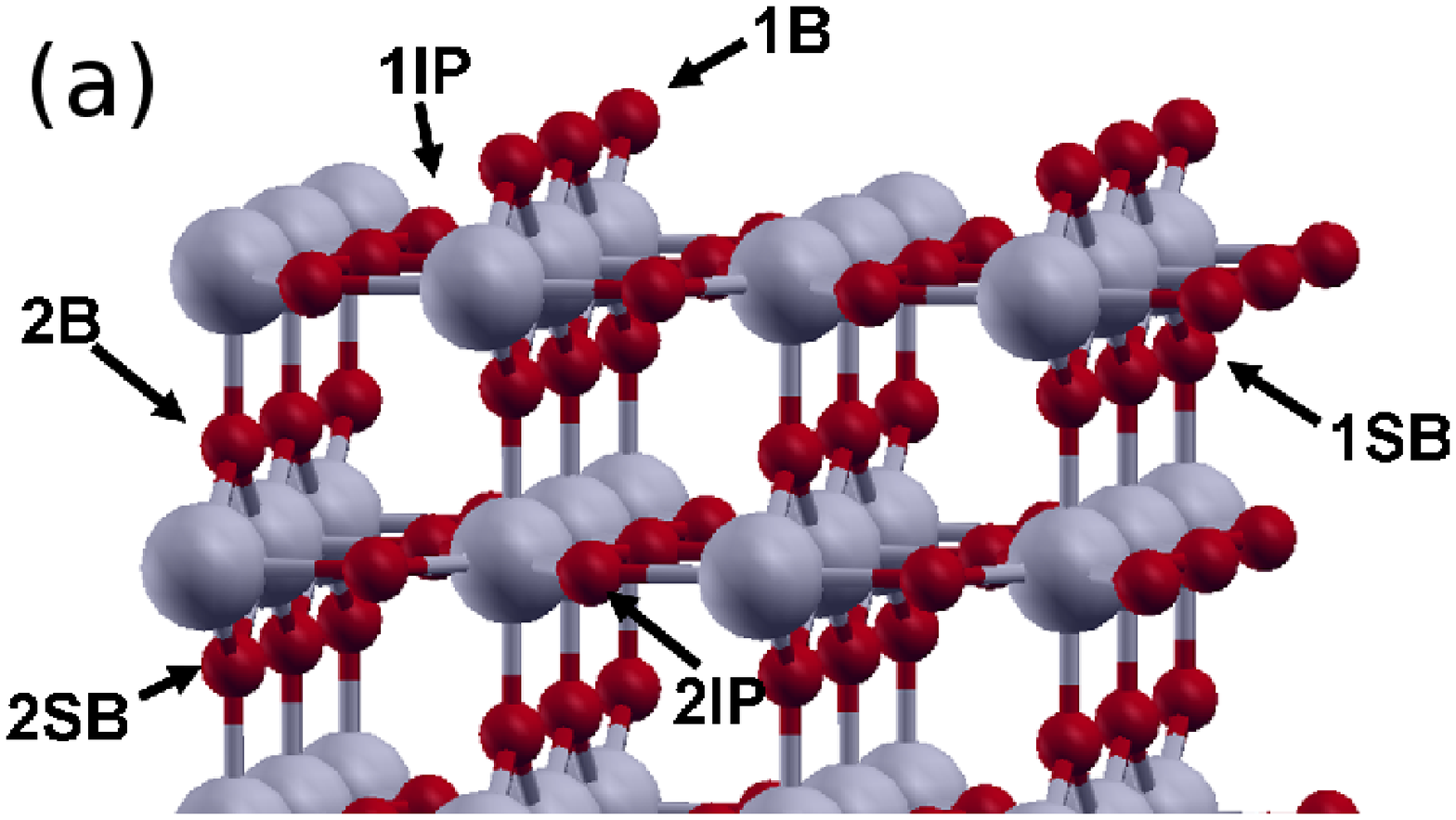}
\includegraphics[width=.4\textwidth]{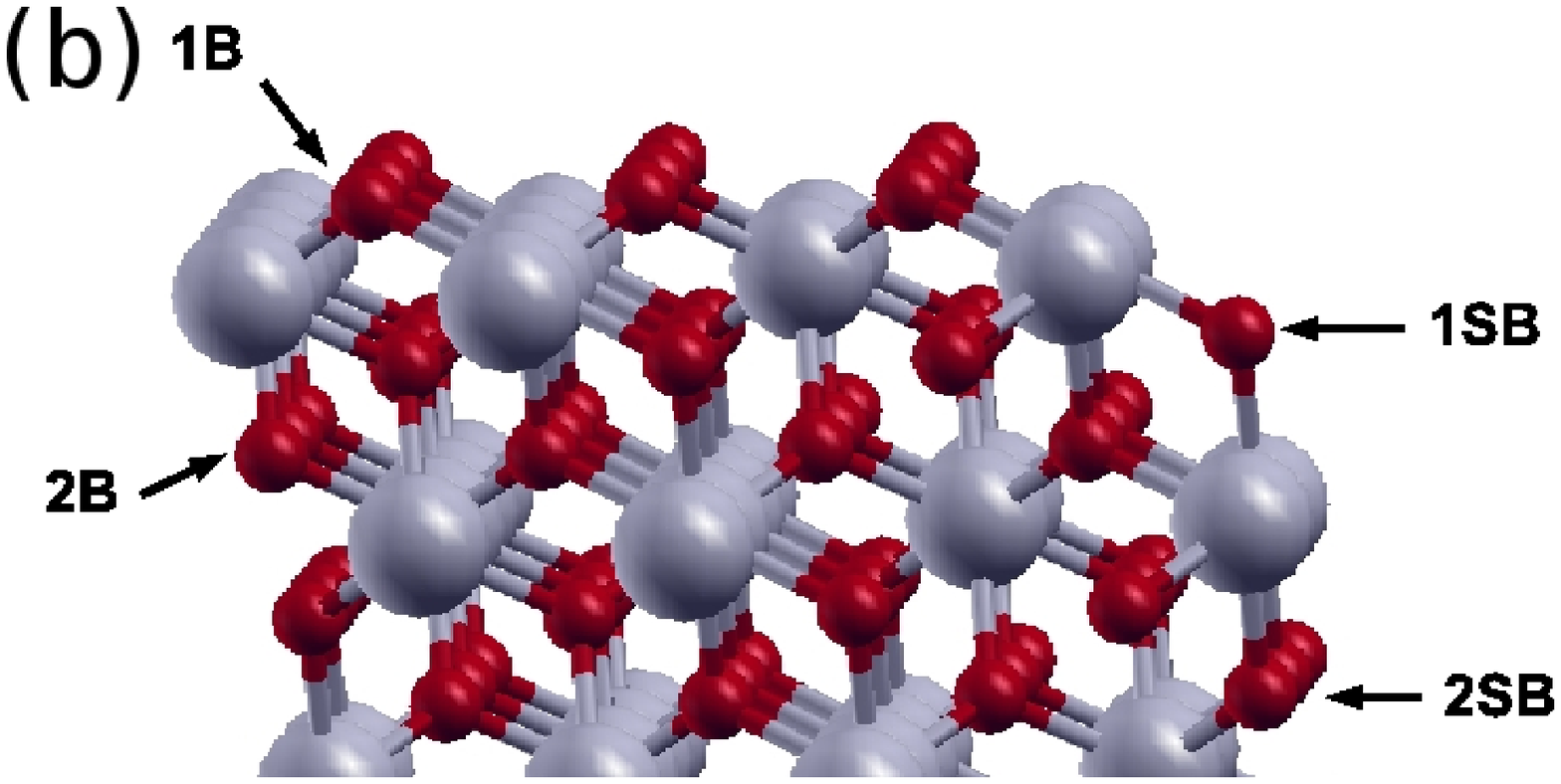}
\caption{\label{fig:fig1}(Color online) Side view of the (a) SnO$_2 (110)$ and (b) SnO$_2 (101)$
stoichiometric surfaces. The large (small) spheres represent tin
(oxygen) atoms. The oxygen atoms are labeled according to the convention
discussed in the text. }
\end{figure}

The second approach, which is based on atomic pseudopotentials and a plane-wave basis set as implemented
in the \textsc{quantum-espresso}\cite{espresso} package, was employed due to its being very efficient for treating
extended, periodic systems. In this case,
the calculations were performed using the local density approximation (LDA) with
the Perdew-Wang parametrization for the exchange-correlation functional and Bachelet-Hamann-Schl\"uter (for tin) and Vanderbilt ultrasoft (for oxygen)
pseudopotentials to represent the ionic cores. We used an energy cutoff of 30 Ry for the wave
functions and 180 Ry for the charge density. The need of using a higher value of the charge density cutoff is due to the fact that we need an accurate estimation of the electrostatic potentials for the calculation of the electron affinities (see below). 

On the basis of our experience,\cite{cantele,buonocore} different computational schemes converge to comparable results when all the main parameters are carefully set. In the present case, we checked the charge density cutoff energy and the vacuum width parameters, for the plane-wave calculations, while we carefully set the Gaussian basis set by adding ghost orbitals (see below) for the all-electron approach.
Once verified that both methods give reliable results, we used them for different types of calculations.
The \textsc{crystal\footnotesize{06}} package was used to perform calculations of band structures and density of states, thus taking advantage of the accuracy of B3LYP exchange-correlation functional in the single particle energy calculations. On the other hand, the \textsc{quantum-espresso} package was used to perform spatial averages [of potential energy and charge density (see below)] over surface planes, align the surface electrostatic potential with the bulk one, and, therefore, perform calculations of electron affinity variations induced by defects.

The rutile SnO$_2 (110)$ and $(101)$ stoichiometric surfaces, which are shown in
Figs. \ref{fig:fig1}(a) and \ref{fig:fig1}(b), respectively, have been
modeled with a five layer (seven layer) slab in the case of \textsc{\textsc{crystal\footnotesize{06}}} (\textsc{quantum-espresso}) calculations.
The slab has been studied as an isolated structure (two-dimensional periodic structure) with the all-electron \textsc{\textsc{crystal\footnotesize{06}}} code, whereas a three-dimensional supercell has been considered in using \textsc{quantum-espresso}.
In the latter case, we have checked that a vacuum width of 6 \AA{} between the slab replicas is large
enough for obtaining converged results. 
For the sampling of the Brillouin zone, we used a 3x3 Monkhorst mesh (and a 6x6 secondary mesh for evaluating the density matrix\cite{crystal}) for the geometrical optimization of the defected structures. Instead, we used a 6x6 Monkhorst mesh (12x12 secondary mesh) for the band structure and density of states calculations. We checked that finer meshes do not change the results.

Defected surfaces have been obtained by removing the oxygen atoms in the positions
labeled in Fig. \ref{fig:fig1}.
The outermost atomic plane of the stoichiometric SnO$_2 (110)$ surface [panel (a)] of Fig. \ref{fig:fig1} is characterized by
rows of oxygen in bridging sites between pairs of tin atoms. The atomic plane lying just below
is made of alternating rows of oxygen and tin atoms. The next plane is again made of
only oxygen, which is again arranged along rows.\cite{manassidis95} The O atoms belonging to the three atomic
planes just mentioned will be referred to as bridging, in-plane, and sub-bridging, respectively.
Bridging, in-plane, and sub-bridging O atoms form a primitive structural unit along the direction perpendicular to the surface, which will be referred to in the following as a surface layer.\cite{kris}
To label these atoms, we use a number for the layer depth (1 is the outermost
layer) and a symbol for the position within the layer: bridging (B),
in-plane (IP), and sub-bridging (SB).
In the same way, the stoichiometric SnO$_2 (101)$ surface shown in panel (b) of Fig. \ref{fig:fig1}  is characterized by rows of bridging (sub-bridging) oxygen atoms, located
above (below) planes of tin atoms.
The stoichiometric surfaces are always favored over other kinds of reconstructions
when samples are treated in oxygen-rich atmosphere.\cite{diebold,bergermayer}
In the present paper, we are interested in analyzing how oxygen vacancies modify the properties of a
stoichiometric surface.
We calculate the vacancy formation energy as
$E_{\text{vf}}=E_{\text{vac}}+\frac{1}{2}E_{\text{O}_{2}}-E_{\text{st}}$,
where $E_{\text{vac}}$, $E_{\text{O}_2}$, and
$E_{\text{st}}$ are, respectively, the total energy of the slab containing the
vacancy, that of an isolated spin polarized O$_{2}$ molecule, and that of the stoichiometric
slab. The present expression is equivalent to Eq. (2) of Ref. \onlinecite{zunger}, when the particular case of a neutral oxygen vacancy is considered.

The geometries of all the systems considered in this work have been optimized by fully relaxing
the atomic positions until the residual forces were less than $10^{-3}$ Ry / Bohr. 
We conclude this section by pointing out that
the inclusion of the spin polarization does not modify our results.
Indeed, we checked that, at variance with titanium dioxide (where in the presence of 
oxygen vacancies, a triplet ground state may arise),
tin dioxide surfaces always show a spin singlet ground state which is 
fully justifying the use of spin unpolarized calculations.

\section{Bulk phase}
\label{sec:bulk}
In Table \ref{tab:bulk}, the calculated lattice parameters, energy gap, and conduction
band effective masses of bulk rutile SnO$_2$
are compared to the experimental data, and show good agreement.
The theoretical results, which are obtained with the all-electron B3LYP calculations, 
show deviations from the experimental data of only about 1.5\% for the structural
properties and less than 3\% for the energy gap.
\begin{table}
\caption{\label{tab:bulk}The lattice parameters ($a$, $c$), the internal coordinate $u$, the energy gap $E_{gap}$, and
the $c$-axis parallel and perpendicular components of the conduction band effective masses 
of cassiterite (bulk rutile SnO$_2$). Theoretical results are obtained from
the B3LYP, all-electron calculation. Experimental data are reported for comparison.
The lattice parameters $a$ and $c$ are in \AA{}, the
energy gap is in eV, and the masses are in free-electron-mass units.}

\begin{ruledtabular}
\begin{tabular}{l|l l l l l l}
  & $a$     & $c$ & $u$ &  $E_{gap}$ & $m_\parallel$ & $m_\perp$ \\
\hline
 B3LYP & 4.74 & 3.24  & 0.306 & 3.5 & 0.22 & 0.26 \\
 EXP &  4.74\footnotemark[1]& 3.19\footnotemark[1] & 0.307\footnotemark[1] & 3.6\footnotemark[2] & 0.23\footnotemark[3] & 0.3\footnotemark[3]
\end{tabular}
\end{ruledtabular}
\footnotetext[1]{Reference \onlinecite{baur}.}
\footnotetext[2]{Reference \onlinecite{diebold}.}
\footnotetext[3]{Reference \onlinecite{button}.}
\end{table}

\begin{figure}
 \includegraphics[width=0.4\textwidth]{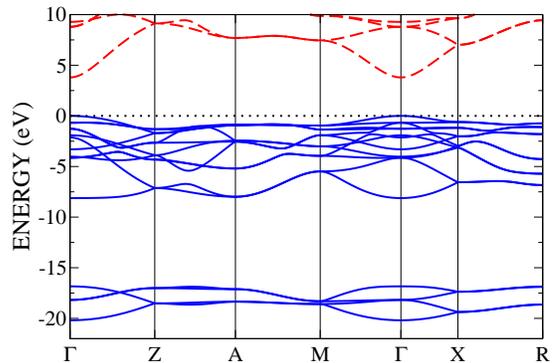}
 \caption{\label{fig:fig2}(Color online) The band structure of bulk SnO$_2$ obtained from the all-electron, B3LYP calculation. Blue solid (red dashed) lines refer to occupied (unoccupied) states. A dotted line indicates the highest occupied level.}
\end{figure}

In Fig. \ref{fig:fig2}, we show the band structure of bulk SnO$_2$. As already known, SnO$_2$ is
a direct band gap semiconductor, with the lowest energy electron-hole
transition occurring at the $\Gamma$ point. It is interesting to compare the
present results with previous band structure calculations.
Former tight binding calculations\cite{godin93} fairly reproduced the lowest bands (located at
15-20 eV under the top valence band) and the overall shape of the conduction band.
Yet, they fail in the calculation of the top valence band, due to a wrong
estimation of the charge transfer toward the oxygen atoms. This failure
is critical in the presence of an oxygen vacancy, mostly if any
geometrical reconstruction at the vacancy site is neglected.\cite{themlin90,manassidis95}
Other interesting results emerge from
a comparison with the GGA density-functional calculations of Ref.
\onlinecite{maki01}. In that paper, M\"aki-Jaskari and Rantala\cite{maki01} showed that ultrasoft pseudopotentials (USP) provide
more accurate results than their norm-conserving (PSP)
counterparts, in the energy range close to the top valence band.
We confirm their conclusion since the shape of
the present B3LYP band structure (see Fig. \ref{fig:fig2}) is in good agreement with the USP calculation within the full valence band spectrum.

\begin{figure}[b]
 \includegraphics[width=0.4\textwidth]{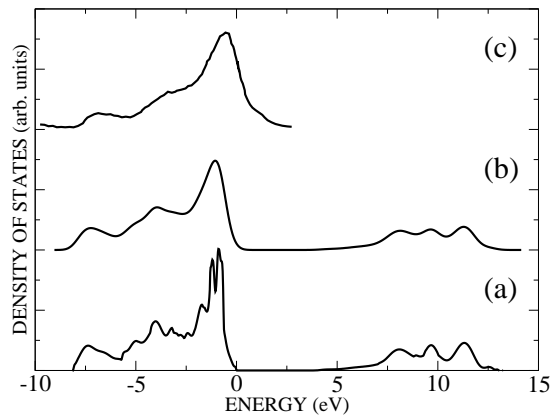}
 \caption{\label{fig:fig3}DOS of bulk tin dioxide.
(a) Present B3LYP calculation. The top of the valence band has been chosen as a reference
energy. (b) The DOS is convoluted with a Gaussian curve whose half width at
half maximum is 0.3 eV. (c) Experimental data taken from Ref.
\onlinecite{themlin90}.}
\end{figure}

In Fig. \ref{fig:fig3}, we show the calculated and measured density of states (DOS) of bulk SnO$_2$. The theoretical result [curve (a)] has also been broadened using 
a Gaussian distribution with a half width at half maximum of 0.3 eV [curve (b)]
to facilitate the comparison with the experimental\cite{themlin90} data
[curve (c)]. Figure \ref{fig:fig3} shows good agreement between theory and
experiment. In particular, we find a valence band width of
8.1 eV in good agreement with both experimental data (7.5 eV reported in Ref. \onlinecite{themlin90})
and previous first-principles calculations (7.9 eV and 8.8 eV reported in Ref. \onlinecite{maki01} using PSP and USP, respectively).

\begin{figure}
  \includegraphics[width=0.45\textwidth]{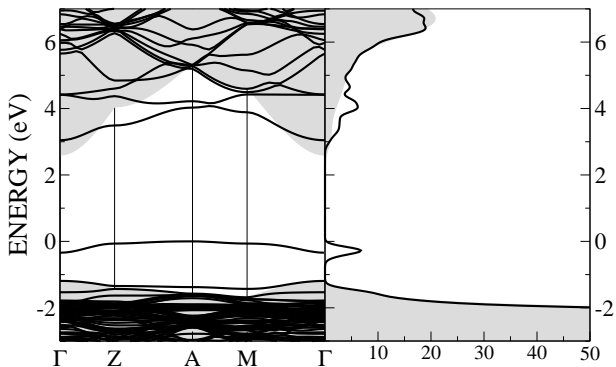}
  \caption{\label{fig:fig4}Band structure (left panel) and DOS (right panel) of
bulk SnO$_2$ with an oxygen vacancy in a bridging site. A 2x2x3
supercell has been used in the calculation. The
shaded regions represent the projected band structure
and DOS of the undefected structure.}
\end{figure}

Before discussing the results concerning the surface band structures with and without oxygen vacancies, it is worth addressing their role in bulk SnO$_2$.
Isolated vacancies have been modeled using a 2x2x3 (almost cubic) supercell,
which contains 12 primitive unit cells. From each supercell, we remove a bridging oxygen atom, 
so that the final structure results in a vacancy-vacancy distance of about 9.5
\AA{}. We computed a vacancy formation energy of 3.42 eV for this configuration.
Figure \ref{fig:fig4} shows the band structure (left panel) and the DOS
(right panel) in the near-gap region for the defected structure (lines).
The shaded regions represent the projected band structure and the DOS of the ideal crystal, respectively.
All the energy levels located at negative (positive) energies
are occupied (unoccupied) at a temperature of 0 $^{\circ}$K. 
The uppermost occupied state of the defected structure is taken as reference energy.
The comparison between
the pure and the defected crystals leads to important conclusions. First, vacancies 
are responsible for the appearance of an almost flat occupied band
which is located about 1 eV above the top valence band of the ideal crystal.
The corresponding peak in the DOS is well visible in the right panel of Fig.
\ref{fig:fig4}, which shows a narrow well-defined distribution.
Interesting results also emerge for the unoccupied states.
Indeed, Fig. \ref{fig:fig4} evidences the appearance of
states related to the oxygen vacancy and resonating with the conduction band. 
The coupling between the conduction
band and the resonance level localized at the vacancy site pushes the bottom of
the conduction band at a higher energy. A clear evidence of the effects of the
vacancy on the energy levels emerges from the DOS. The vacancy is
responsible for both the blue-shift of the density of unoccupied states threshold and
the appearance of a strong peak at about 1.1 eV above this threshold.

\section{SnO$_2 (110)$ surface}
\label{sec:110}

As outlined in sec. \ref{sec:comp}, different vacancy sites
(bridging, in-plane, and sub-bridging) are possible for the
SnO$_2 (110)$ surface. We have considered all these vacancy positions,
at several depths inside the slab. In all cases, a $2\times1$ surface reconstruction 
(the $x$ and $y$ axes have been chosen along
the $[001]$ and $[1\bar{1}0]$ crystallographic axes, respectively) is considered.

\begin{figure}
 \includegraphics[width=0.4\textwidth]{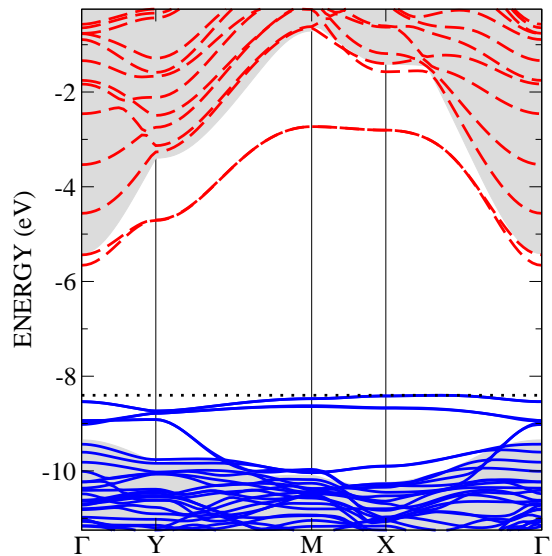}
 \caption{\label{fig:fig5}(Color online) Band structure of the SnO$_2 (110)$ stoichiometric surface. The reference energy is the vacuum level. The last occupied state is at -8.4 eV and is indicated with a dotted line. Occupied (unoccupied) bands are reported as blue solid (red dashed) lines. The shaded region corresponds to the projected bulk band structure. The alignment of the projected band structure is fixed to match the $1s$ core levels of the Sn atoms in bulklike positions. The bands are calculated using the B3LYP scheme.}
\end{figure}

The band structure of the stoichiometric (undefected) tin dioxide surface
is reported in Fig. \ref{fig:fig5}.
The reference energy is the vacuum level.
The comparison with the bulk SnO$_2$ band structure evidences that surface
bands show up inside the forbidden band gap. The projected DOS (PDOS)
onto atomic orbitals shows that the top valence band
mostly consists of $p$ states localized on the external bridging (1B) oxygen
atoms. The bottom of the conduction band mainly consists of $s$ orbitals
localized on the in-plane tin atoms. 
The formation of the intragap surface states leads to a surface band gap
much smaller than the bulk one. We find that the stoichiometric surface has
a band gap at the $\Gamma$ point of $2.41$ eV.
Figure \ref{fig:fig5} is in qualitative agreement with previous theoretical
calculations,\cite{maki01} with the difference that the present approach gives
reliable results for the conduction band energies. An indication of the good
quality of the present calculation can be found in the position of the lowest
conduction band with respect to the vacuum level. We find a value of $-5.7$ eV
[$-5.5$ eV for the SnO$_2 (101)$ surface], whereas, secondary electron cutoff measurements in photoemission gave a work function value (defined as the difference between the vacuum and the Fermi level energies) of $5.7\pm0.2$ eV for the stoichiometric SnO$_2 (101)$ surface.\cite{batzill04} According to the literature, the Fermi level lies just below the lowest conduction level,\cite{diebold} (tin dioxide behaves as a $n$-type material), and thus, the agreement with the present calculation is good.

\begin{figure}
 \includegraphics[width=0.45\textwidth]{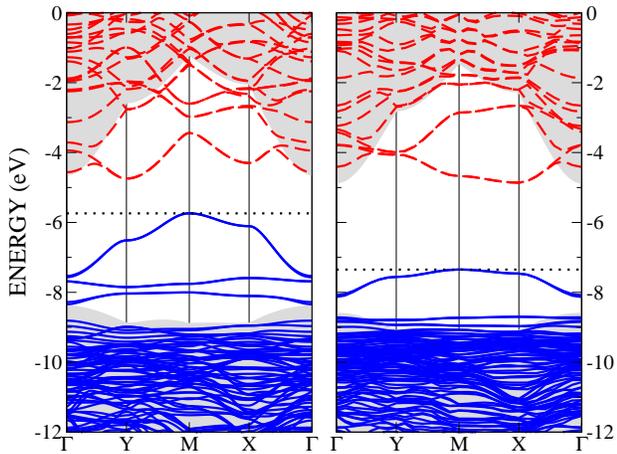}
 \caption{\label{fig:fig6}(Color online) Band structure of the SnO$_2 (110)$ surface with an
oxygen vacancy at the outermost bridging (1B) (left panel) or in-plane (1IP)
(right panel) positions. The reference energy is the vacuum level. The highest occupied states lie at -5.74 eV (left panel) and -7.35 eV (right panel) and are indicated by dotted lines. Occupied (unoccupied) bands are reported in blue solid (red dashed) color. The shaded region corresponds to the projected bulk band structure. The alignment of the projected band structure is fixed to match the $1s$ core levels of the Sn atoms in bulklike positions. The bands are calculated using the B3LYP scheme.}
\end{figure}

In Fig. \ref{fig:fig6}, we show the band structures of a defected SnO$_2 (110)$
surface, where either a bridging (1B, left panel) or an in-plane (1IP, right panel)
oxygen atom has been removed. There are substantial differences with respect to
the stoichiometric surface, especially concerning the intragap states.
As a result of the removal of the most exposed
(1B) bridging oxygen atoms, a dispersed band arises within the forbidden
gap. The trend found here is in agreement with the experimental literature, reporting that, upon reducing a stoichiometric surface, energy levels extend from the top bulk valence band up
to about 0.5 eV below the conduction band minimum.\cite{cox}
We found a different picture when the vacancy lies in the subsurface positions.
In this case, the levels within the gap more closely resemble those of bulk SnO$_2$,
with a little dispersed band placed at about 1 eV above the top bulk valence band, as
shown in the right panel of Fig. \ref{fig:fig6}.

\begin{figure}
 \includegraphics[width=0.45\textwidth]{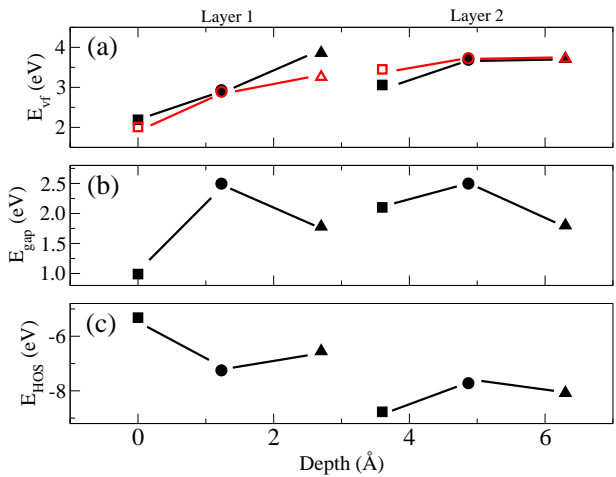}
 \caption{\label{fig:fig7}(Color online) (a) Vacancy formation energy, (b) energy gap,  and (c)
highest occupied state position with respect to the vacuum level.
Bridging (squares), in-plane (circles), and sub-bridging (triangles) oxygen
vacancies are considered.
Black full and red empty symbols represent calculations performed using either the all-electron B3LYP \textsc{\textsc{crystal\footnotesize{06}}} package or
the pseudopotential plane-wave LDA \textsc{quantum-espresso} package. The lines are guides for the eyes.}
\end{figure}

In Fig. \ref{fig:fig7}, we report the vacancy formation energy
[panel (a)], the energy gap  [panel (b)], and the highest occupied state (HOS)
binding energy calculated with respect to the vacuum level [panel (c)], upon changing
the depth of the oxygen vacancy inside the slab. Squares,
circles, and triangles correspond to the bridging, in-plane and sub-bridging oxygen
vacancies, respectively.
The black full symbols show calculations performed using the all-electron Gaussian based \textsc{\textsc{crystal\footnotesize{06}}} package, with the B3LYP
exchange correlation functional.
As a comparison, the red empty symbols represent calculations performed using the plane-wave pseudopotential \textsc{quantum-espresso}
package, with a LDA for the exchange correlation functional.
Interesting results can be inferred from this figure.
First, the vacancy formation energy [Fig. \ref{fig:fig7}(a)] changes with the depth
going toward the bulk limit of 3.42 eV for sufficiently deep vacancy sites.
Although there is some scattering in the data, there is always a monotonic
increasing behavior of the curve at a fixed vacancy position [compare
1B-2B, 1IP-2IP, and 1SB-2SB in Fig. \ref{fig:fig7}(a)]. In
other words, for a given position, the vacancy formation energy either increases
or remains constant. This result is at variance with that found in the case of
titanium dioxide, where a more oscillating behavior comes out.\cite{kris1}

The results on tin dioxide vacancy formation energies already available in the literature span a wide range
of values. There is a strong dependence on the vacancy type, depth, and
concentration, but also on the computational method used for the
calculations. For the outermost bridging oxygen vacancy (1B in Fig. \ref{fig:fig1}), the plane-wave
calculations of Oviedo and Gillan\cite{oviedo1} and M\"aki-Jaskari and Rantala\cite{maki02}
gave values of  2.6 eV and 3.0 eV, respectively. On the other hand, the localized orbital
calculations of Sensato \textit{et al.}\cite{sensato02} lead to a value of 4 eV for the formation energy
of a fully reduced surface, in which all bridging atoms had been removed.
A possible source of discrepancy among the methods can be found in the erroneous evaluation, within a Gaussian basis set, of the charge density in the space region of the missing oxygen,\cite{scorza} due to the non negligible ionicity of tin dioxide.
In order to fix this error, we added three Gaussian type functions to the
original basis set centered at the vacancy site (ghost orbitals).\cite{scorza}
We verified that the addition of ghost orbitals reduces the vacancy formation energy (it is $0.5$ eV for the 1B vacancy), which leads
to a better agreement with the plane-wave results.
Another important source of discrepancy lies in the use of either PSPs or USPs in the plane-wave approach, which causes strong variations of the vacancy formation energies.\cite{maki01}
The data reported in Fig. \ref{fig:fig7} show the overall fair agreement that we obtained for the vacancy formation energy using either the all-electron B3LYP Gaussian basis set or the pseudopotential plane-wave LDA calculations.

Figure \ref{fig:fig7}(b) shows the energy gap  $E_{\text{gap}}$ as a function of the vacancy depth.
$E_{\text{gap}}$ strongly depends on the hybridization of atomic orbitals
surrounding the vacancy, which explains the dependence on the vacancy
site. An interesting point is that, with the exception of the more exposed (1B)
oxygen whose removal leads to the lowest value of the energy gap  (due to the
formation of many states within the gap), for each vacancy type
there is a negligible dependence on the depth. There is a gap for bridging, in-plane, and
sub-bridging vacancies. Since such a behavior seems very promising for technological applications,
further calculations are being performed on deeper vacancy sites.
Figure \ref{fig:fig7}(c) shows the HOS binding energy
with respect to the vacuum level. There is an overall reduction of the HOS
energy upon increasing the depth at which the vacancy is formed.
This reduction of the HOS is in line with the electron affinity increase (see below).
Unfortunately, in the literature, there is a lack of experimental measurements of the HOS energy versus the vacancy depth.

\begin{figure}
 \includegraphics[width=0.45\textwidth]{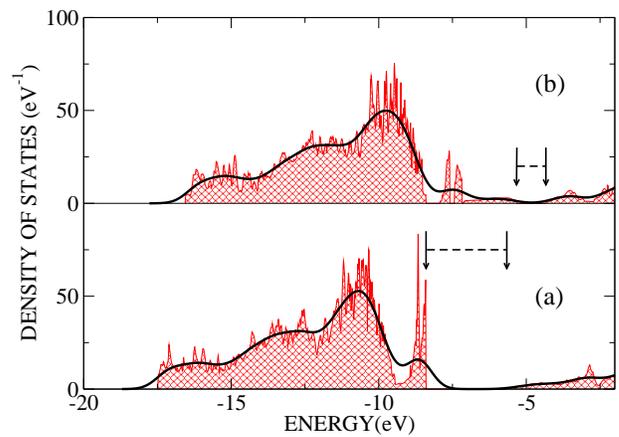}
 \caption{\label{fig:fig8}(Color online) DOS of (a) the stoichiometric and (b) the defected (1B O vacancy)
SnO$_2 (110)$ surface. The (smoother) thick lines are a convolution of the DOS
with a Gaussian broadening of 0.4 eV. We indicate with the shaded region the original (nonconvoluted) DOS.
The highest occupied and lowest unoccupied states are reported in the figure using vertical arrows, whereas
the forbidden energy window is indicated by a horizontal dashed line.}
\end{figure}

In Fig. \ref{fig:fig8}, we show the DOS of both the stoichiometric surface
[panel (a)] and of the surface with a bridging (1B) oxygen vacancy [panel (b)].
The reference energy is the vacuum level. As already mentioned, the surface reduction (e.g., after
sputtering) leads to the formation of intragap
surface states,\cite{cox,themlin90} but the oxidized (stoichiometric) surface
shows surface states as well (see Fig. \ref{fig:fig5}).
This explains the experimental finding that the annealing process in
air does not fully remove the intragap states.
We find that the stoichiometric surface is featured by occupied intragap states
at about 1 eV above the top bulk valence band. This is in agreement with previous
calculations\cite{maki01,manassidis95} and experiments.\cite{themlin90}
Upon reduction of the surface, a continuum of occupied states which covers a wide energy
range within the bulk band gap is experimentally reported.\cite{cox}
This clearly emerges from Fig. \ref{fig:fig8}(b). The main
peak above the bulk valence band is lower in value than the corresponding peak of
the stoichiometric surface. At the same time, a wider energy range within the bulk gap
is spanned by the occupied states. In the figure, the highest occupied and lowest unoccupied states
are indicated with arrows and the forbidden energy window between those states is reported
using a dashed line. The reduction in the energy gap due to the presence of the vacancy is
clearly seen.
From Fig. \ref{fig:fig8}, we also observe the shift to higher energies of the whole bulk
spectrum of the defected with respect to the stoichiometric surface. This is
in agreement with the decrease in work function by about 1 eV reported in
the experimental literature.\cite{batzill04,batzill06}

\section{SnO$_2 (101)$ surface}
\label{sec:101}

For the study of oxygen vacancies at the SnO$_2 (101)$ surface, we have
considered a $\sqrt{2}\times\sqrt{2}$ reconstruction (double cell, i.e.,  the area is
doubled with respect to the unit cell of the bulk geometry). We studied the
vacancy formed by removing a top bridging oxygen atom from each unit cell 
with a vacancy-vacancy distance of 7.4 \AA{}.

\begin{figure}
   \includegraphics[width=0.45\textwidth]{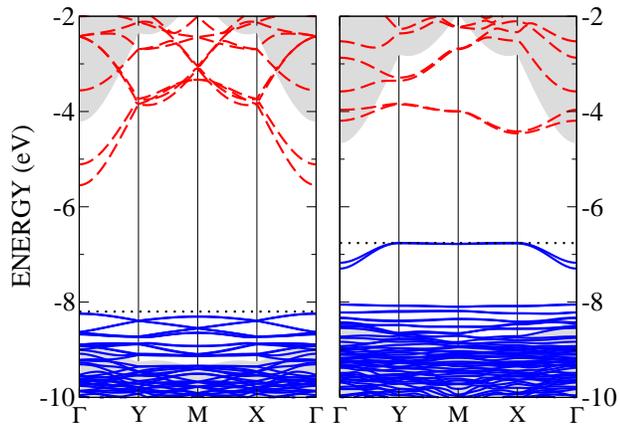}
 \caption{\label{fig:fig9}(Color online) Band structure of the stoichiometric (left) and
defected (bridging oxygen vacancy, right) SnO$_2 (101)$ surfaces. The last occupied states are at -8.2 eV and -6.76 eV, respectively, and are indicated by a dotted line. Occupied (unoccupied) bands are reported in blue solid (red dashed) color. The shaded region corresponds to the projected bulk band structure. The alignment of the projected band structure is fixed to match the $1s$ core levels of the Sn atoms in bulklike positions. The bands are calculated using the B3LYP scheme.}
\end{figure}

In Fig. \ref{fig:fig9}, the band structures of the stoichiometric (left panel)
and the defected (right panel) surfaces are shown. In order to compare both 
band structures, a $\sqrt{2}\times\sqrt{2}$ reconstruction is considered for
the stoichiometric surface as well. Similarly to what is discussed in the case of the
SnO$_2 (110)$ surface, the presence of the surface leads to the
formation of surface bands within the bulk band gap. The stoichiometric surface has
a direct gap at $\Gamma$, with an energy gap of 2.7 eV. The presence of the
oxygen vacancy leads to the formation of a doubly degenerate occupied energy
level, which is well separated from all the other bands. This level shows a parabolic
dispersion near $\Gamma$, but it is almost flat elsewhere inside the first Brillouin zone..
From the PDOS analysis, we found that the corresponding states are well
localized around the vacancy site, which explains in this way the lack of
dispersion. The oxygen vacancy is also responsible for a level resonant with
the conduction band. At variance with the stoichiometric surface, the bottom of
the conduction band for the defected surface lies at the $X$ point.

The first allowed transition for the SnO$_2 (101)$ surface when a top
bridging oxygen vacancy is formed is at the $X$ point. Interestingly, the 
electron and hole states involved in this direct-gap recombination
are quite localized around the vacancy site. This localization
leads to a strong oscillator strength and may explain the 
photoluminescence activity of tin dioxide nanowires.\cite{luo,lettieri}

\section{Electron affinity}
\label{sec:affinity}

Electron affinity (EA) variations induced by the presence of vacancies have been
studied for oxygen vacancies located on the first two layers of the SnO$_2 (110)$
surface [see Fig. \ref{fig:fig1}(a)]. This part of the work has been
done using the \textsc{quantum-espresso} code.
The EA can be easily calculated once the vacuum level has been
determined from the self-consistent electrostatic potential.\cite{ivo}
The results are
collected in Table \ref{tab:ea}. In this table, we give, for each vacancy type,
the depth of the removed oxygen calculated from the top bridging oxygen.
It is of interest to see how the different vacancies contribute to the EA.

\begin{figure}
 \includegraphics[width=0.45\textwidth]{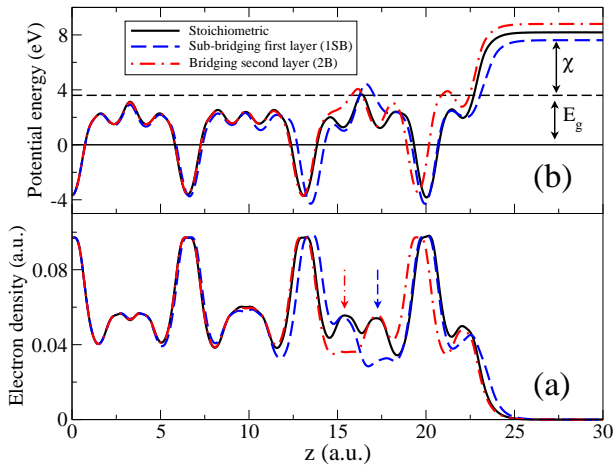}
 \caption{\label{fig:fig10}(Color online) Planar average of the self-consistent electrostatic
potential energy (upper panel) and electron density (lower panel) for the
stoichiometric SnO$_2 (110)$ surface. The ideal surface is compared with the
defected one (1SB and 2B oxygen vacancies). All the potentials calculated for the
studied configurations are aligned with the bulk potential in the middle layer
of the slab. The energy reference is taken on the top of the bulk SnO$_2$ valence
state. The electron affinity $\chi$ is defined as the difference between the
vacuum level and the bulk conduction state energy.
The position of the oxygen vacancy within the slab is indicated with an arrow in the lower panel.
The calculations have been done using a pseudopotential plane-wave approach.}
\end{figure}

The top panel of Fig. \ref{fig:fig10} shows the planar
average of the electrostatic potential. The averages are taken on 
planes parallel to the surface
(the $z$ coordinate lies along the surface normal and $z$=0 
corresponds to the slab center). The potential is shown for the stoichiometric case
and for the 1SB and 2B vacancies. 
If we take the stoichiometric surface as a reference, it is seen that the 2B vacancy
induces an EA variation of 0.62 eV, whereas the variation due to the 1SB vacancy 
is of -0.57 eV. This difference in both the amplitude and sign arises from
the different electron charge density and atomic relaxations following
the oxygen removal. To better illustrate this point, we plot in the bottom
panel of Fig. \ref{fig:fig10} the planar averaged electron charge density.
An electron density depletion occurs in correspondence of the
vacancy site, as indicated by the vertical arrows in the figure. Moreover,
the atomic relaxation can be inferred from the shift of the charge
density peaks. The combination of both the
charge depletion and the atomic relaxation gives rise to an electric dipole
variation (that we have calculated as in Ref. \onlinecite{ivo}) $\Delta P=0.27$ a.u. for 1SB and $\Delta
P=-0.29$ a.u. for 2B. The change in the electron affinity due to a change
$\Delta P$ in the surface dipole is $\Delta\chi=-4\pi\Delta P/A$, where $A$ is
the surface unit cell area (all quantities measured in atomic
units).\cite{luth,ivo} Within the surface dipole model, we therefore obtain
$\Delta\chi=-0.57$ eV for the 1SB vacancy and $\Delta\chi=0.61$ eV for the 2B
vacancy with remarkable agreement with the data of Table \ref{tab:ea}.

\begin{table}
 \caption{\label{tab:ea}EA and its variation
with respect to the stoichiometric surface ($\Delta$EA), calculated for the
SnO$_2 (110)$ surface with an oxygen vacancy, whose depth is reported in the
second column. The vacancies are labeled according to Fig.
\ref{fig:fig1}(a). All the energies are measured in eV, and the depth is in \AA{}.}
\begin{ruledtabular}
\begin{tabular}
[c]{ccccc}
Vacancy type & Depth & EA & $\Delta $EA \\
\hline
1B  & 0.0 & 3.69 & -0.90\\
1IP & 1.23  & 4.00 & -0.59\\
1SB & 2.70  & 4.02 & -0.57\\
2B & 3.60 & 5.20 & 0.62\\
2IP & 4.87  & 4.54 & -0.05\\
2SB & 6.30  & 4.91 & 0.33\\
Stoichiometric  &  & 4.59 & 0.0\\
\end{tabular}
\end{ruledtabular}
\end{table}

Few experimental works have been published on the dependence of electron
affinity and work function on the surface oxidation.\cite{cox,szuber} In
particular, in Ref. \onlinecite{cox}, a careful study was done on the
effects of the annealing temperature on the electron affinity and conductivity
of an oxidized (presumably stoichiometric) SnO$_{2} (110)$ surface. Cox \textit{et al.}\cite{cox} found
that upon increasing the annealing temperature, the amount of oxygen lost by
the surface increases. They concluded that for temperatures up to about 700
$^{\circ}$K, only the bridging oxygen (1B in Fig. 1) is removed from the surface
followed, at higher temperatures, by the in-plane oxygen. This trend is in
qualitative agreement with the vacancy formation energies shown in Fig. \ref{fig:fig7}(a). 
The results of Ref. \onlinecite{cox} show that
the variation of electron affinity with respect to the stoichiometric surface
is negative, with a maximum variation of about $0.4-0.5$ eV upon increasing
the annealing temperature. This is compatible with our results in
Table \ref{tab:ea}.  However, it should be stressed that a strict comparison
between our results and the experimental data is made difficult by both the
uncertainties in the actual surface structure and the computational
difficulties in analyzing a large number of different vacancy configurations.

Finally, we conclude by noticing that the recently reported
decrease in the work function of about $1$ eV upon reducing the surface\cite{batzill04, batzill06} is in
agreement with the results shown in Table \ref{tab:ea}, where the electron
affinity decreases by $0.9$ eV when an outermost oxygen atom is removed from
the stoichiometric surface.

\section{Conclusions}
\label{sec:concl}

By using \textit{ab initio} computational methods, we have studied the formation of oxygen
vacancies at the SnO$_{2} (110)$ and $(101)$ surfaces and their role on the
surface electronic properties.  As expected, the formation energy tends to
increase, when the vacancy moves from the surface to subsurface layers. The energy
gap, instead, is very sensitive to the vacancy geometry (and reconstruction),
but it only weakly depends on the layer depth inside the material. Interesting
results on the alignment of energy levels and the appearance of intragap states
have been discussed in connection with the presence of oxygen
vacancies. 

Some results have also been obtained concerning
the role of oxygen vacancies in bulk tin dioxide.
We have found that oxygen vacancies do not lead to the 
formation of occupied donor levels which are just below the
conduction band. Instead, the present calculations have shown that occupied
levels associated to the vacancy appear at about 1 eV above the top valence
band. 

Another important conclusion of the present work is
that the process of surface reduction, which leads to the removal of the most
external oxygen atoms, pushes the density of states spectra to higher energies,
as already known from experimental results.

Finally, our calculations have shown that oxygen
vacancies may significantly modify the material electron affinity.
Interestingly, subsurface vacancies can either increase or decrease the
electron affinity and this effect can be described in terms of a surface
dipole layer that takes contributions from both the atomic relaxation and
electron charge depletion around the missing oxygen. These findings could help
in understanding the basic physics of the SnO$_{2}$ surfaces.

\begin{acknowledgments}
Financial support from the projects PON S.Co.P.E.
and PON STSS-500 is acknowledged. Calculations were
performed at CINECA-Bologna (``Progetti Supercalcolo 2007'') 
advanced computing facilities.
\end{acknowledgments}


\end{document}